\documentclass[aip,amsmath,amssymb, bm, reprint]{revtex4-1}
\usepackage{graphicx}  
\usepackage{amsmath}
\usepackage{bm}        
\usepackage{amssymb}   
 
\usepackage{float}

\begin{document}
\title{Rotational and translational dynamics in dense fluids of patchy particles}

\author{Susana Mar\'in-Aguilar}
\affiliation{Universit{\'e} Paris-Saclay, CNRS, Laboratoire de Physique des Solides, 91405, Orsay, France}
\author{Henricus H. Wensink}
\affiliation{Universit{\'e} Paris-Saclay, CNRS, Laboratoire de Physique des Solides, 91405, Orsay, France}
\author{Giuseppe Foffi}
\email{giuseppe.foffi@u-psud.fr}
\affiliation{Universit{\'e} Paris-Saclay, CNRS, Laboratoire de Physique des Solides, 91405, Orsay, France}
\author{Frank Smallenburg}
\email{frank.smallenburg@u-psud.fr}
\affiliation{Universit{\'e} Paris-Saclay, CNRS, Laboratoire de Physique des Solides, 91405, Orsay, France}

\date{\today}
\begin{abstract}
We explore the effect  of directionality on rotational and translational relaxation in glassy systems of patchy particles. Using molecular dynamics simulations we analyze the impact of two distinct patch geometries, one that enhance local icosahedral structure and one which does not strongly affect local order. We find that in nearly all investigated cases, rotational relaxation takes place on a much faster time scale than translational relaxation.  By comparing to a simplified dynamical  Monte Carlo model, we illustrate that rotational diffusion can be qualitatively explained as purely local motion within a fixed environment, which is not coupled strongly to the cage-breaking dynamics required for translational relaxation. Nonetheless, icosahedral patch placement has a profound effect on the local structure of the system, resulting in a dramatic slowdown at low temperatures which is strongest at an intermediate ``optimal'' patch size.
\end{abstract}
\maketitle

\section{Introduction}

When we cool down or compress a liquid, the overall dynamics of the particles slow down as they become trapped in cages formed by their neighbors. If the system successfully avoids crystallization, this process eventually leads to a fully arrested state, where relaxation and diffusion are no longer possible on time scales that are relevant to a human observer. In this limit, the liquid has transformed into a glass \cite{ediger1996supercooled}. The nature of this slowdown, and in particular its relation to changes in the structure of the liquid, has been the subject of intensive debate over the last decades \cite{Berthier2011,hunter2012physics,stillinger2013glass,royall2015role}.

One of the characteristic features of glasses is the emergence of dynamical heterogeneities \cite{lombardo2006computational, mishra2015shape}: for sufficiently cold or dense systems, the dynamics in the system vary strongly in space, such that some particles are significantly more mobile than others. Inspired by this, a number of studies have linked the dynamical slowdown of glasses to variations in their local structure \cite{doye2003favored, miracle2007structural, tarjus2005frustration, royall2015role, fragiadakis2017role}. For example, it has been demonstrated that slow dynamics are paired with an increasing emergence of particles arranged in locally favored structures: local packings that are energetically or entropically highly favorable \cite{doye2003favored, tanaka2005relationship} and, thus, survive for a long time \cite{malins2013identification}. Additionally, recent work using machine learning techniques have demonstrated that the likely mobility of a particle can be predicted purely on its local structure \cite{schoenholz2016structural}.

If local structures play an important role in determining the dynamics of glassy systems, a natural question to ask is whether we can then control dynamics by deliberately modifying the local geometry of particle packings. Recently, we showed \cite{shortpaper} that the dynamics of a model glass former (a binary sphere mixture) can be tuned by including directional (``patchy'') interactions that favor specific local environments. When the favored environment had icosahedral symmetry, this induced a dramatic slowdown in the relaxation of the system, as compared to all other tested geometries. Hence, for these models, icosahedral structures are an ideal tool for controlling dynamical slowdown.

While patchy interactions are a great tool for controlling the local structure of glassy systems, their overall impact on the system goes beyond merely changing the local structure. Despite still being spherical in shape, the patchy particles are inherently anisotropic, and hence perform both translational and rotational motion. As a result, not only the position of a particle can be caged by neighbors, but also its orientation. In most glassy materials, both of these types of motion slow down as the glass transition is approached. However, rotations and translations do not necessarily slow down at the same rate, and -- depending on the interactions between the particles -- may become arrested at significantly different temperatures or packing fractions \cite{dzugutov2002decoupling, edmond2012decoupling, mazza2007connection, chong2002structural, chong2005evidence, mishra2015shape, roos2016coupling}.
In the case of dense systems of spherical patchy particles, an interesting contrast arises. While the translational motion of a particle is hindered both by the repulsive cores of its neighbors and the attractive bonds it has formed with them, rotational motion is purely the results of the attractive bonds. As a result, it is natural to expect a decoupling between translational and rotational motion, at least when the temperature is high enough for the repulsive forces to dominate. This is in sharp contrast to the glassy behavior of low-valence patchy particles at low densities, where any dynamical slowdown is essentially dominated by the time scale at which the bonding network changes, typically leading to strong glass forming behavior \cite{de2006dynamics, rovigatti2011self, bianchi2011patchy, smallenburg2013liquids, tavares2017dynamics}.

Here, we examine the interplay between rotational and translational relaxation in more detail in dense suspensions of patchy particles. We use extensive molecular dynamics simulations to investigate patchy particles with two distinct patch geometries. The first favours icosahedral local structures, and hence promotes dynamical slowdown, while the second promotes an octahedral symmetry, which has little effect on the dynamics. For the first case, we find an optimal patch size that maximizes the formation of icosahedral clusters, resulting in extremely slow dynamics. Additionally, we show that the rotational relaxation decouples from translational relaxation in nearly all cases. We conclude that while translational dynamics are controlled by collective effects, the rotational relaxation is purely dependent on the local structure.

\section{Methods}
\subsection{Model}
To model our particles, we use a variation on the Kern-Frenkel model \cite{Kern}. The model consists of hard-sphere particles decorated with $n$ patches on their surface. Two particles form an attractive bond when they are within a maximum interaction range of each other, and when the vector that joins their centers of mass passes through a patch on each particle. Hence, the potential energy in the Kern-Frenkel model is as follows:
\begin{equation}
U_{ij}(\mathbf{r}_{ij})=U_{ij}^{\textsc{HS}}(r_{ij})+U_{ij}^{\textsc{SW}}({r}_{ij})f(\mathbf{r}_{ij},\hat{\mathbf{n}}_{\alpha}, \hat{\mathbf{n}}_{\beta}),
\end{equation}
where $r_{ij} = |\mathbf{r}_{ij}|$ is the center-to-center distance between particles $i$ and $j$. Here, $U_{ij}^{\textsc{HS}}$ is the hard-sphere potential:
\begin{equation}
U_{ij}^{\textsc{HS}}= \begin{cases} 
\infty & r_{ij} \leq \sigma_{ij} \\
0 & r_{ij} > \sigma_{ij},
\end{cases}
\end{equation}
with $\sigma_{ij} = (\sigma_i + \sigma_j)/2$ the minimum distance between two particles, and $\sigma_i$ the diameter of particle $i$. Additionally, $U_{ij}^{\textsc{SW}}$ is a square-well potential, given by
\begin{equation}
U_{ij}^{\textsc{SW}}= \begin{cases} 
\epsilon & r_{ij} \leq \sigma_{ij}+\lambda_{ij} \\
0 & r_{ij} > \sigma_{ij}+\lambda_{ij},
\end{cases}
\end{equation}
where we choose the interaction range $\lambda_{ij} = 1.033 \sigma_{ij}$.
Finally, $f(\hat{r}_{ij},\hat{\mathbf{n}}_{\alpha}, \hat{\mathbf{n}}_{\beta})$ specifies the directionality of the interactions: 
\begin{equation}
f(\mathbf{r}_{ij},\hat{\mathbf{n}}_{\alpha}, \hat{\mathbf{n}}_{\beta})=\begin{cases}
1  & \begin{cases}\hat{\mathbf{n}}_{\alpha} \cdot \hat{\mathbf{r}}_{ij} < \cos \theta \text{\,\,and\,\,}
\hat{\mathbf{n}}_{\beta}\cdot \hat{\mathbf{r}}_{ji} < \cos \theta, \\
\text{for any two patches $\alpha$ and $\beta$}
\end{cases}
 \\
0 
\end{cases}
\end{equation}
where $\hat{\mathbf{n}}_{\alpha}$ corresponds to a unit vector that points to patch $\alpha$ on particle $i$, and $\hat{\mathbf{r}}_{ij}= \mathbf{r}_{ij} / r_{ij}$. The angle $\theta$ controls the size of the patches. Note that unlike in the original Kern-Frenkel model, two particles can never form more than one bond, even if patches overlap on the same particle. The bonding energy is restricted to be either $0$ or $\epsilon$ even if the vector that joins two particles passes through two or more patches. this allows to interpolate between a hard-sphere model when $\theta\!=\!0$ and a square-well when the surface is fully covered by the patches. 

\begin{figure}

\includegraphics[width=0.9\linewidth]{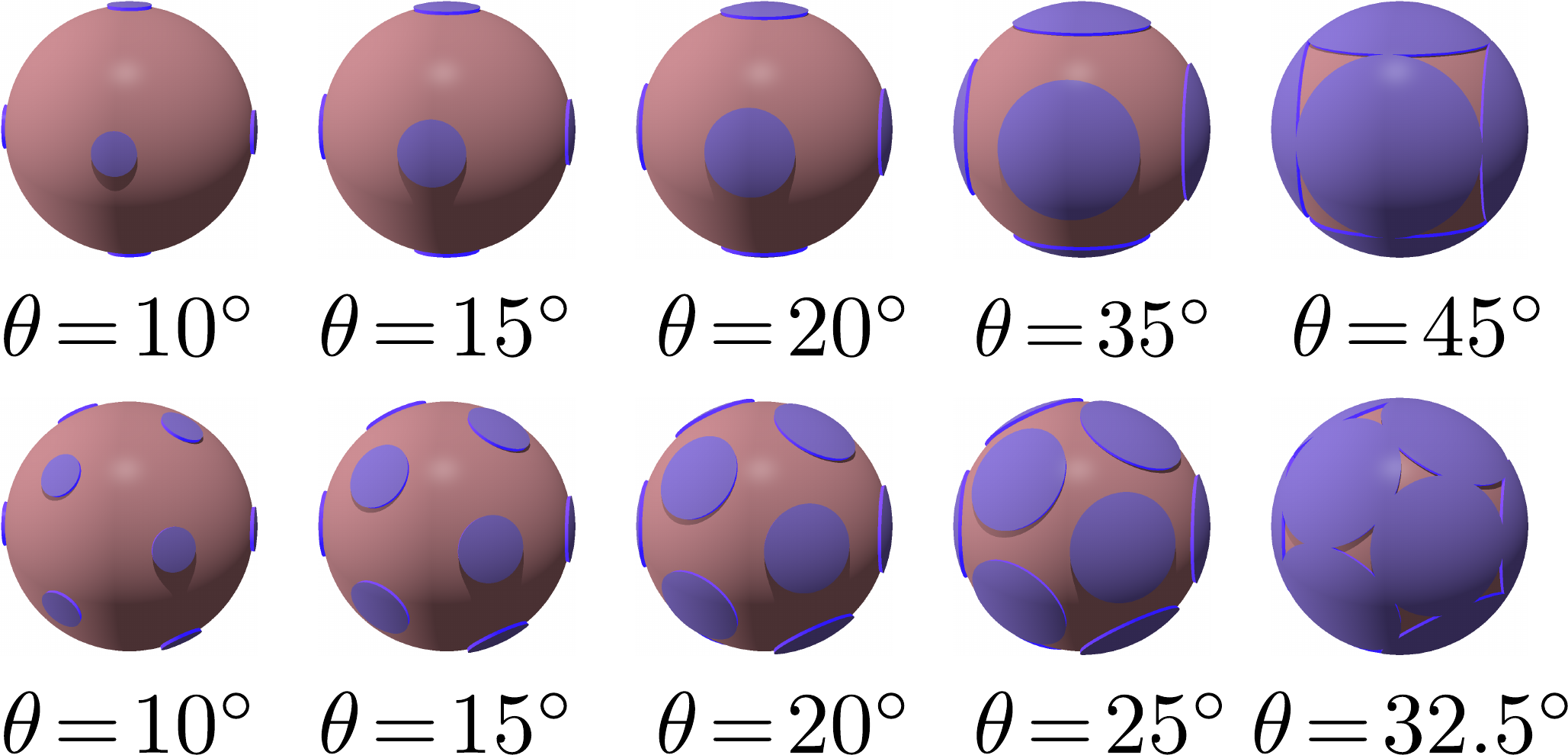} 

\caption{Illustration of the simulation model. The top row shows particles with 6 attractive patches with varying patch sizes determined by the opening angle $\theta$. The bottom row shows the corresponding images for 12 attractive patches.}
\label{fig:model}
    \end{figure}

In this work, we focus on patchy particles with either 6 or 12 patches.
The patches are located in a way that maximizes the minimal distance between the nearest patches. The location of the patches in the 12-patch case corresponds to an icosahedral geometry and for the 6-patch to an octahedral geometry. In Fig.~\ref{fig:model}, we present some schematic cartoons of the patch geometry for various patch sizes. Previous work has shown that the 12-patch case is optimal for enhancing local icosahedral order, while the 6-patch case is an example of a geometry which does not significantly enforce a specific local order \cite{shortpaper}. In order to avoid crystallization, we use a binary mixture of particles with a size ratio of $\sigma_S / \sigma_L = 0.83$, where $S$ and $L$ denote the small and large species, respectively. In all cases, the composition $x = N_L / N = 0.5$, where $N_L$ is the number of large particles and $N$ is the total number of particles. For both investigated patch geometries, this mixture has proven to be extremely effective at avoiding crystallization at all temperatures \cite{shortpaper}.  

The model covers two limiting cases, depending on the patch size $\theta$. It coincides with the hard-sphere limit when $\theta=0$, and the square-well limit when the patches cover the entire surface of the particles with an opening angle $\theta > \theta_F$. It is important to note the existence of an angle $\theta_O$ where the patches start overlapping. For the 12-patch case $\theta_O \simeq 31.72^\circ$ and full coverage is achieved at $\theta_F \simeq =37.37^\circ$. The corresponding angles for the 6-patch case are $\theta_O \simeq45^\circ$ and $\theta_F \simeq54.73^\circ$ respectively. We denote the total percentage of the particle surface area covered by patches with $\chi$. When the patches do not overlap, i.e. $\theta < \theta_O$, 
\begin{equation}
    \chi=\frac{n}{2}(1-\cos \theta), 
\end{equation}
with $n$ the number of patches. For larger patch angles, we calculate the coverage fraction numerically.

\subsection{Event-driven molecular dynamics}

We use event-driven molecular dynamics (EDMD) \cite{hernandez2007discontinuous,smallenburg2013liquids} to simulate the dynamics of our model. The systems are composed of $N=700$ particles and the packing fraction is fixed at $\eta=0.58$. The systems were equilibrated at constant temperature for at least $10^4\tau$ where $\tau = \sqrt{m \sigma_L^2/k_B T}$ is our time unit, $T$ is the temperature,  $m$ is the mass of a particle and $k_B$ is Boltzmann's constant. All particles have equal mass and moment of inertia. After equilibration, we simulate the dynamics at constant energy for at least $10^5\tau$ to perform our measurements. In order to ensure that each pair of particles can form at most a single bond between them, the EDMD code only predicts bonding events between particles that are not already bonded. Additionally, during a bond-breaking event the code checks whether there is another combination of patches through which the pair of particles could still be considered bonded. If so, the particles are now considered to be bonded via the new pair of patches, and their motion is unaffected.

In the following, we present the dynamical results for the large particles only. The small particles present qualitatively the same behavior.

\section{Results}
\subsection{Translational relaxation times}

In the supercooled regime a dynamical arrest arises as consequence of caging effects. The time it takes to break these cages, and hence the relaxation time of the supercooled fluid, is strongly linked to the local structure of the system \cite{royall2015role,  shintani2006frustration, tanaka2003roles, miracle2007structural, doye2003favored}. In order to explore the dependence of the relaxation time on the patch size, we measure the time-dependent density correlation function (also known as the intermediate scattering function): 
\begin{equation}
F(q,t)=\frac{\left\langle \rho(-\mathbf{q},t)\rho(\mathbf{q},0)\right\rangle}{\left\langle \rho(-\mathbf{q},0)\rho(\mathbf{q},0)\right\rangle},
\end{equation}
where $\rho(\mathbf{q},t) = \sum_j \exp(-i\mathbf{q} \mathbf{r}_j(t)) $ is the Fourier transform of the density whit $\mathbf{r}_j(t)$ the position of particle $j$.

\begin{figure}[!ht]
\begin{center}
\includegraphics[width=0.9\linewidth]{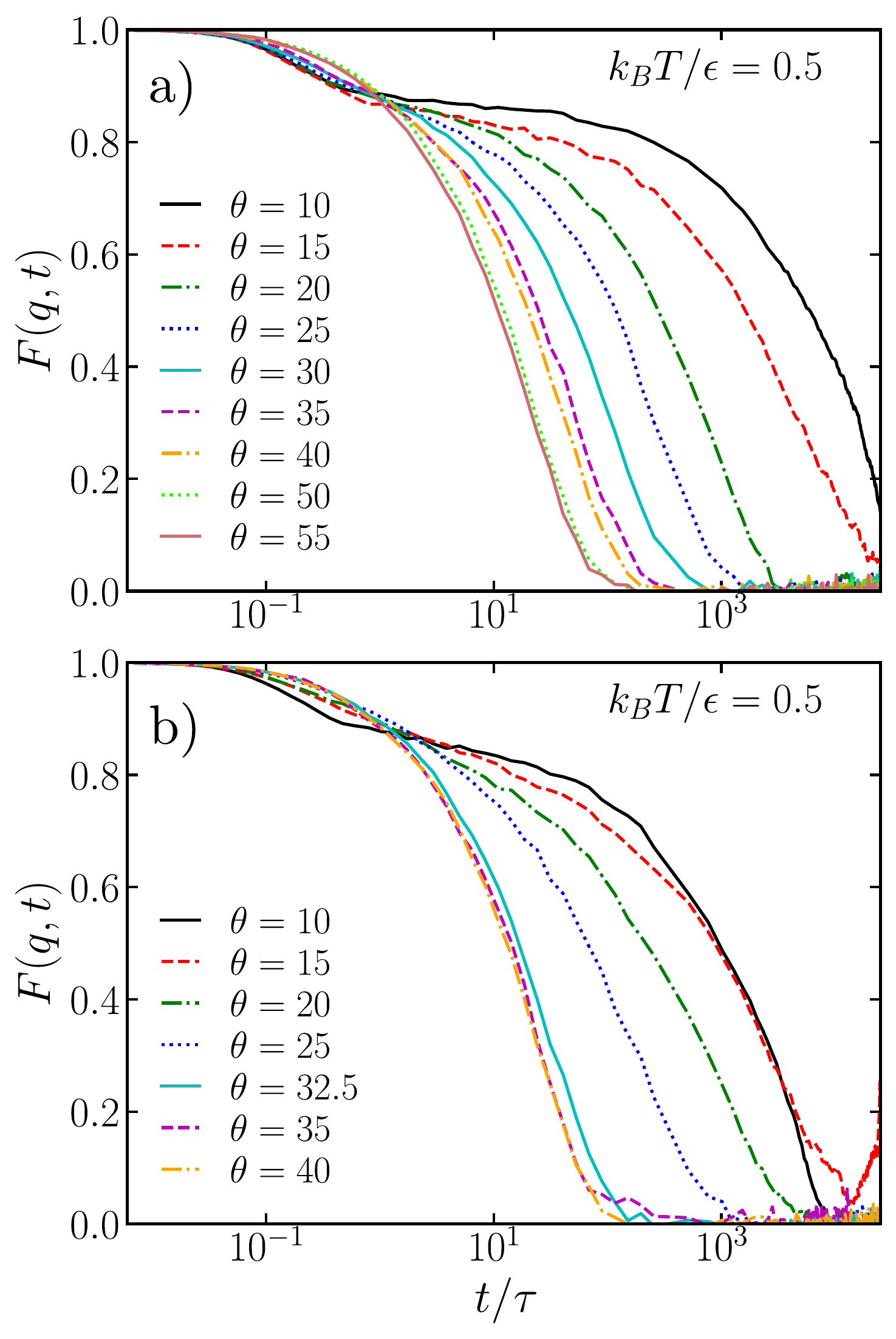}
\end{center}
\caption{Density correlators for the a) 6-patch system and  b) 12-patch system. Both of them of measured for the wavelength corresponding to the first peak of the structure factor.}
\label{fig:dens_correlators}
\end{figure}

In Fig.~\ref{fig:dens_correlators}, we show the correlators corresponding to a relatively high temperature $k_BT/\epsilon \!= 0.5$, for both 6 and 12 patches and a range of different patch sizes. The correlators show significantly different relaxation behavior depending on the size of the patch. For both patch geometries, the smaller angles correspond to much slower relaxation. This can be understood by considering the reentrant glass transition that occurs in these systems, where stronger attractions induce a transition from a repulsive
glass to a fluid, and then to an attractive glass \cite{shortpaper, Sciortino}. At $k_BT/\epsilon\!=\!0.5$, we are in the regime where stronger attractions induce faster dynamics. As larger patches make it easier for the particles to form attractive bonds, these attractions free up additional space in the system, facilitating cage-breaking. As a result, the slowest systems are those with the lowest patch coverage fraction $\chi$, where we see a clear plateau in the correlators. This also explains why the 6-patch systems are generally slower than the 12-patch systems, as they have lower patch coverage for the same patch size.

\begin{figure}[!ht]
\begin{center}
\includegraphics[width=0.9\linewidth]{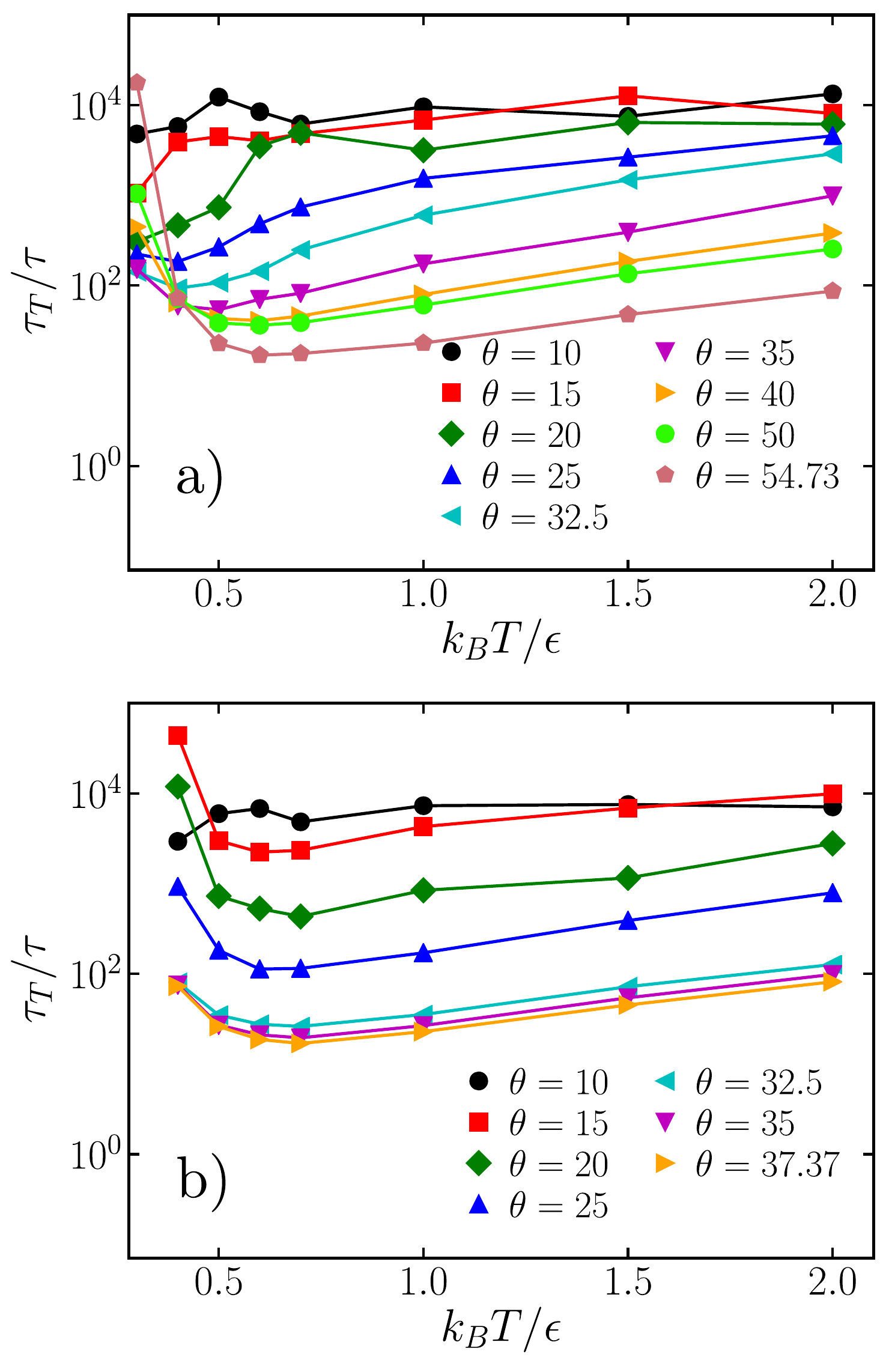}
\end{center}
\caption{Translational relaxation times $\tau_T$ as a function of $k_BT/\epsilon$ for the a) 6-patch and b) 12-patch system.}
\label{fig:tau-T}
\end{figure}

In order to measure the translational relaxation time ($\tau_T$), we fit  the density correlators with a stretched exponential and define the time $\tau_T$ as the time where this fit decays to 0.3. The relaxation shows a clear dependence on the patch size $\theta$. Figure~\ref{fig:tau-T} shows the $\tau_T$ for the 6-patch and 12-patch system as a function of temperature. Note that for most patch sizes, we again see the reentrant dynamical behavior observed in Ref. \onlinecite{shortpaper}, which shifts to lower temperatures as the angle $\theta$ decreases. Overall, we again see that smaller angles lead to stronger dynamical arrest, as we approach the hard-sphere limit.

\subsection{Rotational relaxation times}

Due to the addition of direction-dependent interactions the relaxation behavior of the rotational degrees of freedom is now a point of interest. In order to explore the rotational dynamics, we calculate the decay of the rotational correlation function \cite{de2001viscous}:
\begin{equation}
    C_n(t)=\frac{1}{N}\sum_{j}^N P_n\left(\hat{\mathbf{x}}(t)\cdot\hat{\mathbf{x}}(0)\right)
\end{equation}
 where $P_n$ is the Legendre polynomial of $n$-th degree and $\hat{\mathbf{x}}$ is a fixed unit vector which rotates along with the particle\footnote{  It should be noted here that in our study of the rotational correlations, we deliberately do not account for the symmetry of the particle. In other words, the rotational correlation function simply checks whether an average particle has rotated with respect to its original orientation, even if the new orientation has (different) patches pointing in the same directions as in the original one. This implies we are looking at the ability of the particle to rotate, rather than find a new configuration that is fully independent of its starting orientation. If we were to take into account the particle symmetry, this would likely result in rotational correlation times that are on the time scale of the lifetime of the cages, since the preferred set of orientations for the central particle will adapt itself to the surrounding cage. As this choice would give us information about the (translational) dynamics of the cages, rather than about the true rotational freedom of the particles, we choose here not to take particle symmetry into account when measuring rotations.}. In a rotationally diffusive system that satisfies the Debye equation, these Legendre polynomials have an exponential decay in time \citep{Bitsanis,Kirchhoff1996}.

\begin{figure}
\begin{center}
\includegraphics[width=0.9\linewidth]{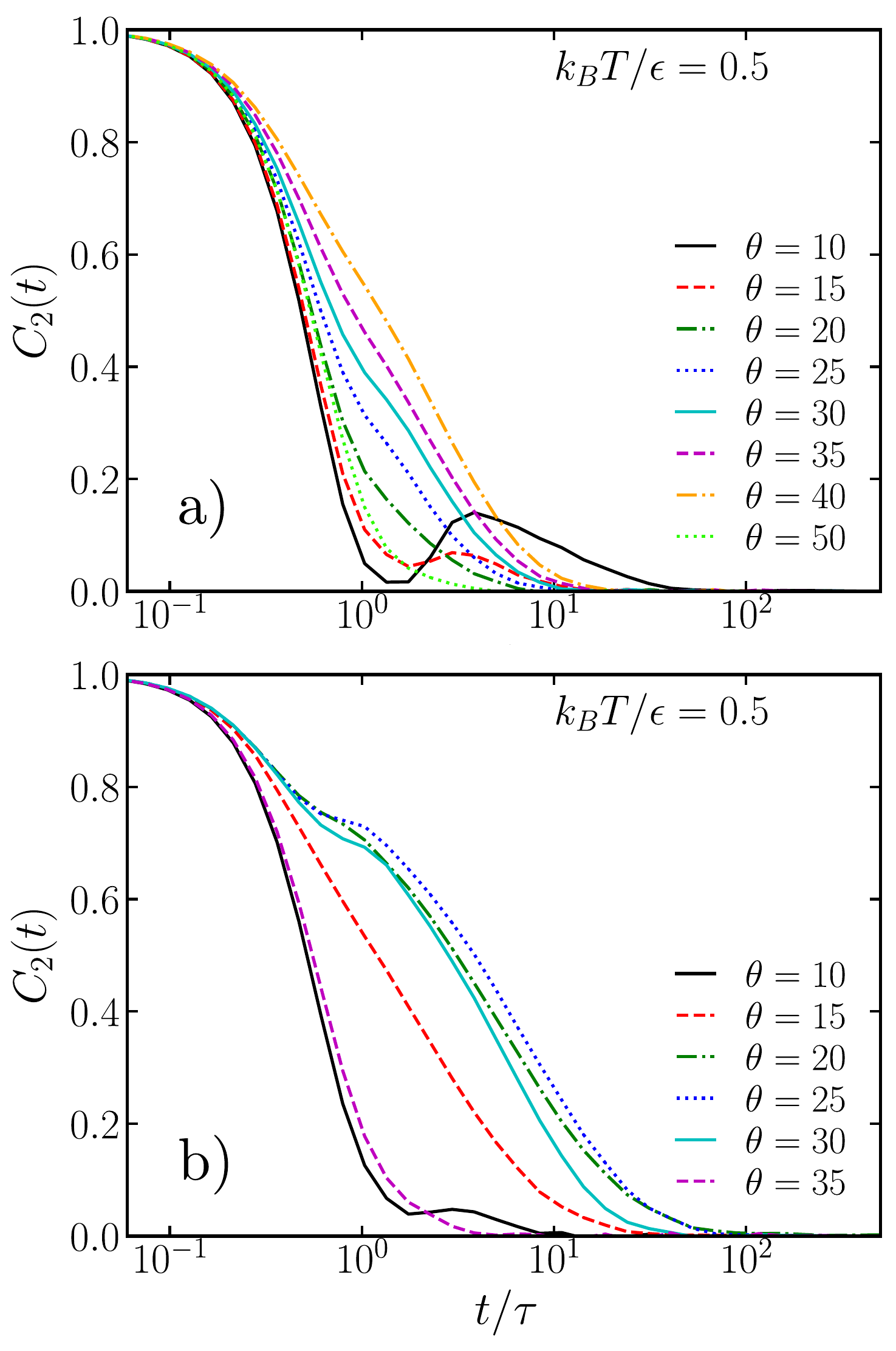}
\end{center}
\caption{Rotational correlators for the a) 6-patch system and  b) 12-patch system.}
\label{fig:rot_correlators}
\end{figure}

Here we take $\hat{x}$ as one of the vectors pointing to a specific patch, such that the dot product $\hat{\mathbf{x}}(t)\cdot\hat{\mathbf{x}}(0)$ is calculated between the initial direction of the patch and the rotated direction after a time $t$. We show in Fig.~\ref{fig:rot_correlators} the  decay of the $C_2$ found in our patchy systems.  As we can see from the figures the rotational correlation functions do not decay exponentially, indicating that the rotations do not satisfy the Debye rotational diffusion equation. At small angles, the rotational correlation function first decays rapidly, and then shows a sudden increase. Note that we see the same behavior for larger $\theta$, but at higher temperatures. We attribute this behavior to the presence of free spinners: particles which rotate around the same axis for a long period of time. As the only interactions that affect the rotation of the particles are the patchy attractions, free spinners occur when the system forms very few bonds, such that patch-patch collisions are rare. Naturally, this predominantly happens when either the patches are very small, or the temperature is high. In order to further clarify this aspect a comparison between a non interacting system with free-spinners is shown in the Supplemental Information (SI).
It is interesting to note that for the 12-patch system, we observe a plateau in the rotational correlation function for intermediate values of $\theta$. This plateau represents the slowing down of rotational dynamics due to bonds with neighboring particles. 
For small angles, bonds are rare, and hence the rotations are not likely blocked, while for sufficiently large angles, the large patches allow a large degree of rotational freedom even in the presence of many bonds. This extra freedom will make it easier to rotate significantly while breaking only a single bond. This results in a reentrant behavior in the rotational relaxation time, where the relaxation is slowest for intermediate $\theta \simeq 25^\circ$. Note that for $\theta = 35^\circ$, the patches overlap, such that 
orientational rearrangements can occur without bond breaking. As one might expect, this plateau becomes more pronounced at lower temperatures. Additionally, at sufficiently low temperatures the plateau also arises in the six-patch system, but mostly for large (non-overlapping) values of $\theta$. Plots of the rotational correlators for other temperatures are shown in the SI.

\begin{figure}[!ht]
\begin{center}
\includegraphics[width=0.9\linewidth]{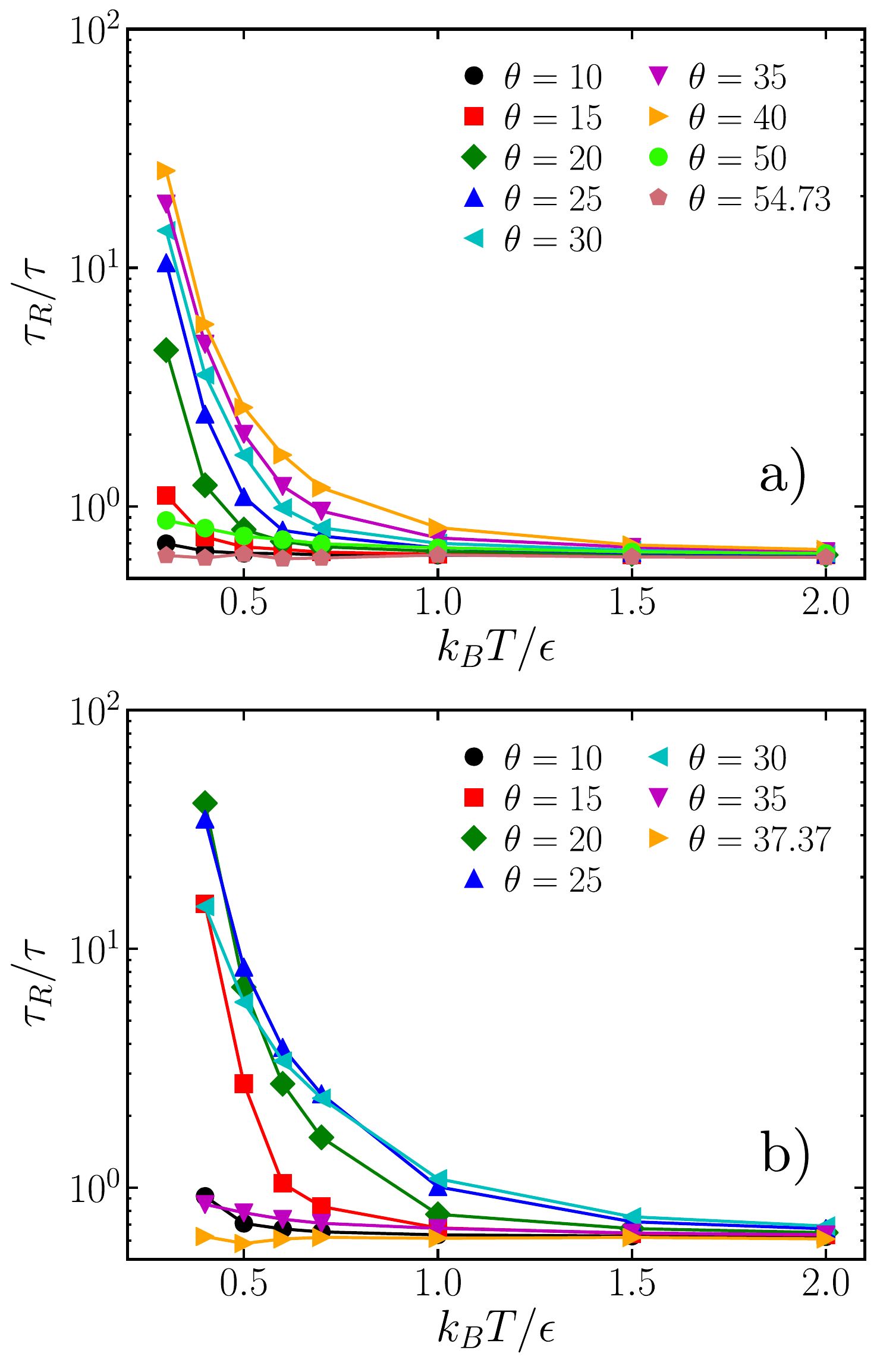}
\end{center}
\caption{Rotational relaxation times $\tau_R$ as function of $k_BT/\epsilon$ for a) the 6-patch system and b) the 12-patch system.}
\label{fig:tau-R_temp}
\end{figure}

In order to measure the rotational relaxation time $\tau_R$, we carefully fit a stretched exponential, avoiding the regime where free spinners give a strong contribution, and define $\tau_R$ as the time at which the correlation function decays to a value of 0.3. In Fig.~\ref{fig:tau-R_temp} we show  $\tau_R$ as a function of temperature for both patch geometries. At low temperatures the rotational relaxation slows down more than one order of magnitude in comparison to higher temperatures. However, a comparison to Fig.~\ref{fig:tau-T} shows that for all investigated systems, the rotational relaxation is significantly faster than the translational relaxation time. With the exception of the lowest investigated temperatures, $\tau_R$ is at least an order of magnitude smaller than $\tau_T$. Moreover, the temperature-dependence of $\tau_R$ shows a stark contrast to that of $\tau_T$: there is no reentrance in the rotation relaxation as a function of temperature. Instead, since only the attractive patchy interactions are capable of slowing down the rotations, the rotational dynamics slow down monotonically as the temperature is lowered. Together, these observations show that the two time scales are essentially decoupled unless the temperature is extremely low. This is in contrast to the expected behavior in fluids outside of the glassy regime, where the Stokes-Einstein and Stokes-Einstein-Debye relations linearly link the rotational and translational diffusivities of a particle via the viscosity of the surrounding fluid \cite{mazza2007connection}. As both of these relationships tend to break down in the glassy regime \cite{dzugutov2002decoupling, edmond2012decoupling, mazza2007connection, chong2002structural, chong2005evidence, mishra2015shape, roos2016coupling}, the qualitatively different behavior between the two modes of motion is not unexpected here. However, it is interesting to observe that even at extremely low temperatures, where most particles are strongly bonded, the particles still manage to rotate almost freely on relatively short time scales.

\begin{figure}[!ht]
\begin{center}
\includegraphics[width=0.9\linewidth]{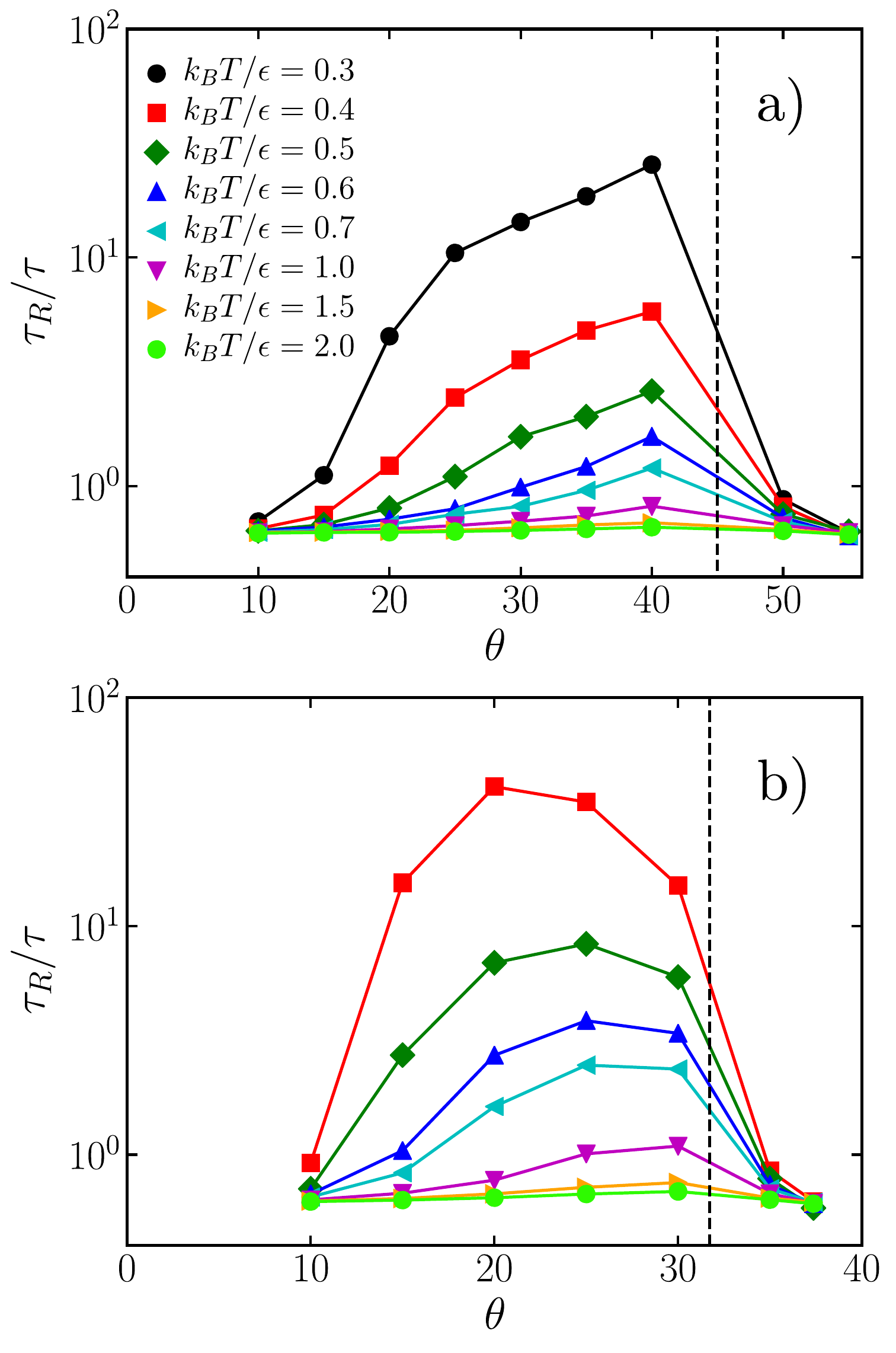}
\end{center}
\caption{Rotational relaxation times $\tau_R$ as a function of the patch opening angle of the system  a) $n=6$ and b) $n=12$. Dashed lines correspond to the overlap angle of each.}
\label{fig:tau_angle}
\end{figure}

In order to investigate the rotational relaxation in more detail, we plot in Fig.~\ref{fig:tau_angle} the dependence of $\tau_R$ on $\theta$ for both systems. The dashed vertical lines show the overlap angle $\theta_O$. Interestingly, there is an angle where the relaxation time is maximized for each patch geometry, but they are qualitatively different. The relation between the opening angle $\theta$ and the relaxation is not straightforward, for the 6-patch case, the slowest rotational dynamics are found just below the overlapping angle: $\tau_R$ increases essentially monotonically until the patches overlap, and then drops sharply as the particle are now much more free to rotate. In contrast, for the 12-patch case, the optimal angle is found well below the overlapping angle, around a value of $\theta \simeq 20^\circ - 25^\circ$ at low temperatures. The 6-patch behavior is the more intuitive behavior: at lower temperatures the particles spend more time bonded to the same patch and hence keep their orientation over longer time scales. Moreover, larger patches lead to more bonds, and similarly slow down rotational dynamics. The only exception to this shows up for $\theta >\theta_O$, where the patches overlap and hence the particle is free to rotate without breaking bonds, resulting in faster relaxation. For the 12-patch case, we see essentially the same behavior for high temperatures, but observe a shift of the maximum to lower $\theta$ when the system is cooled down. As at these low temperatures, the 12-patch system induces strong icosahedral ordering \cite{shortpaper}, this change in dynamical behavior might be attributed to strong changes in the local structural environment of the particles.

In all cases, the rotational correlation function decays at least an order of magnitude faster than the translational correlation function. An important difference between the two is the fact that rotational relaxation can be achieved purely via a local rearrangement of bonds, allowing a particle to rotate within its local cage. This is in contrast to the collective motion that is required for a particle to break out of its cage and diffuse translationally.

\begin{figure}[!ht]
\begin{center}
\includegraphics[width=0.9\linewidth]{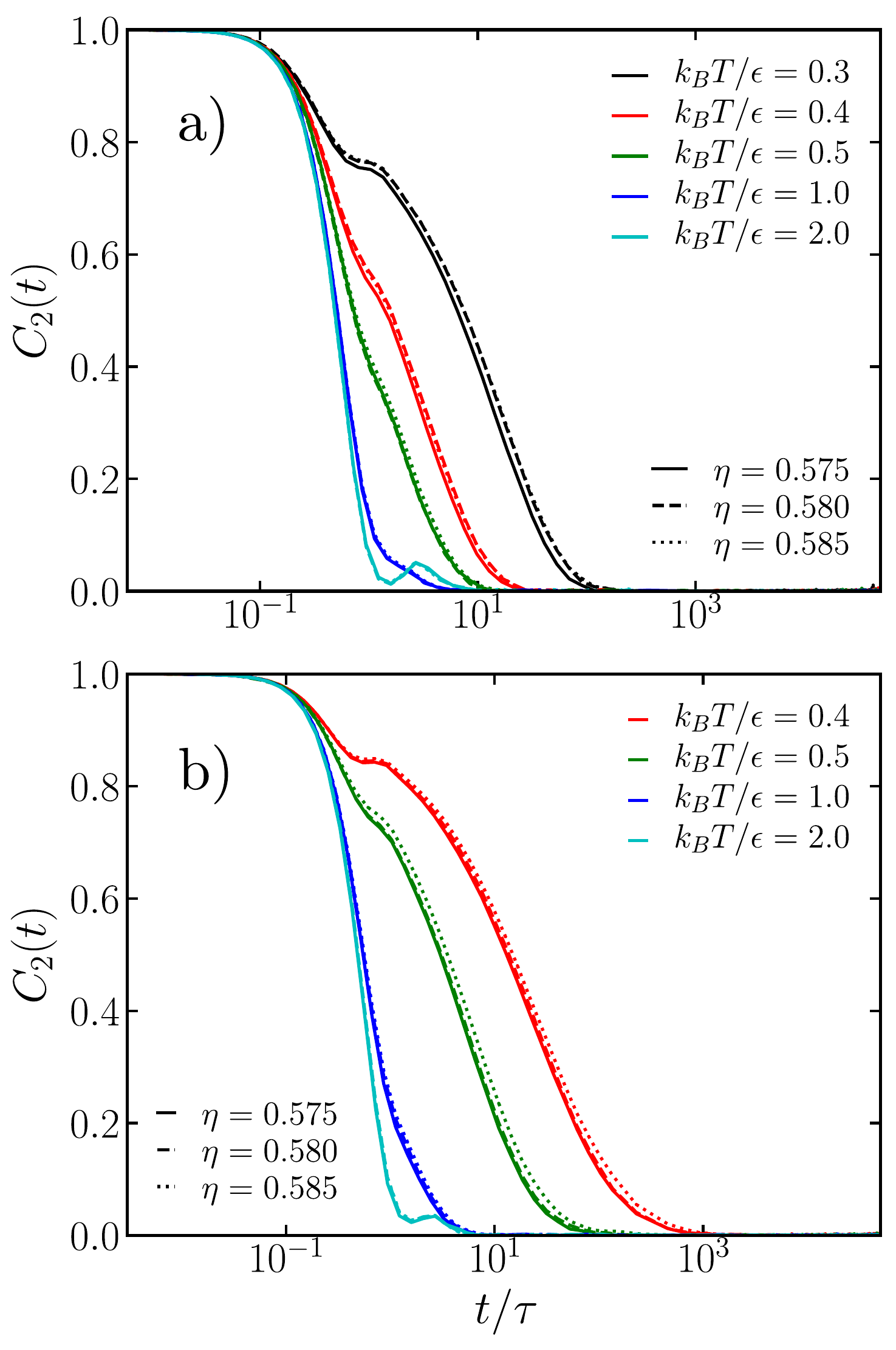}
\end{center}
\caption{Decay of the second Legendre polynomial of a system of a) 6 patches and b) 12 patches. Rotational correlation  with different packing fractions. Dashed lines correspond to a packing fraction $\eta=0.575$, continuous lines to $\eta=0.58$ and dotted lined to $\eta=0.585$.
}
\label{fig:rot_dens}
\end{figure}

Finally, we examine the effect of packing fraction on the   rotational behavior, by performing simulations at two different packing fractions, $\eta=0.585$ and $\eta=0.575$, with an opening angle corresponding to a $\chi=40\%$ and covering a wide range of temperatures. The results show that rotational relaxation is essentially density-independent in the glassy regime, see Fig.~\ref{fig:rot_dens}.

\subsection{Dynamical Monte Carlo}
In the interest of proving that rotational relaxation is mainly due to local motion we develop a simple methodology to sample local rotational dynamics. Specifically, we use a simple model to approximate the rotational relaxation of each particle in its local cage.

Starting from equilibrated configurations from the molecular simulations, we fix the positions of all particles except a single one. We then sample the rotations in one local cage by performing a Monte Carlo simulation on only the chosen particle. During each simulation step, the particle is moved with small rotational and translational displacements, while keeping all other particles fixed. The step size for both types of moves is chosen to be sufficiently small to ensure a high acceptance rate ($\gtrsim 90\%$), such that the dynamics of the particles resemble a Brownian motion trajectory and avoids the possibility of ``jumping'' between two distinct bonded configurations without a change in energy.
In this sense, the present algorithm is dynamical Monte Carlo (DMC) scheme~\cite{sanz2010dynamic}. After $10^6$ steps, the chosen  particle is returned to its initial configuration and the same procedure is done with the next particle. In this manner, we sample the average rotational relaxation of all particles in their respective cages, under the assumption that this relaxation is an entirely local process.

\begin{figure}[!ht]
\begin{center}
\includegraphics[width=0.9\linewidth]{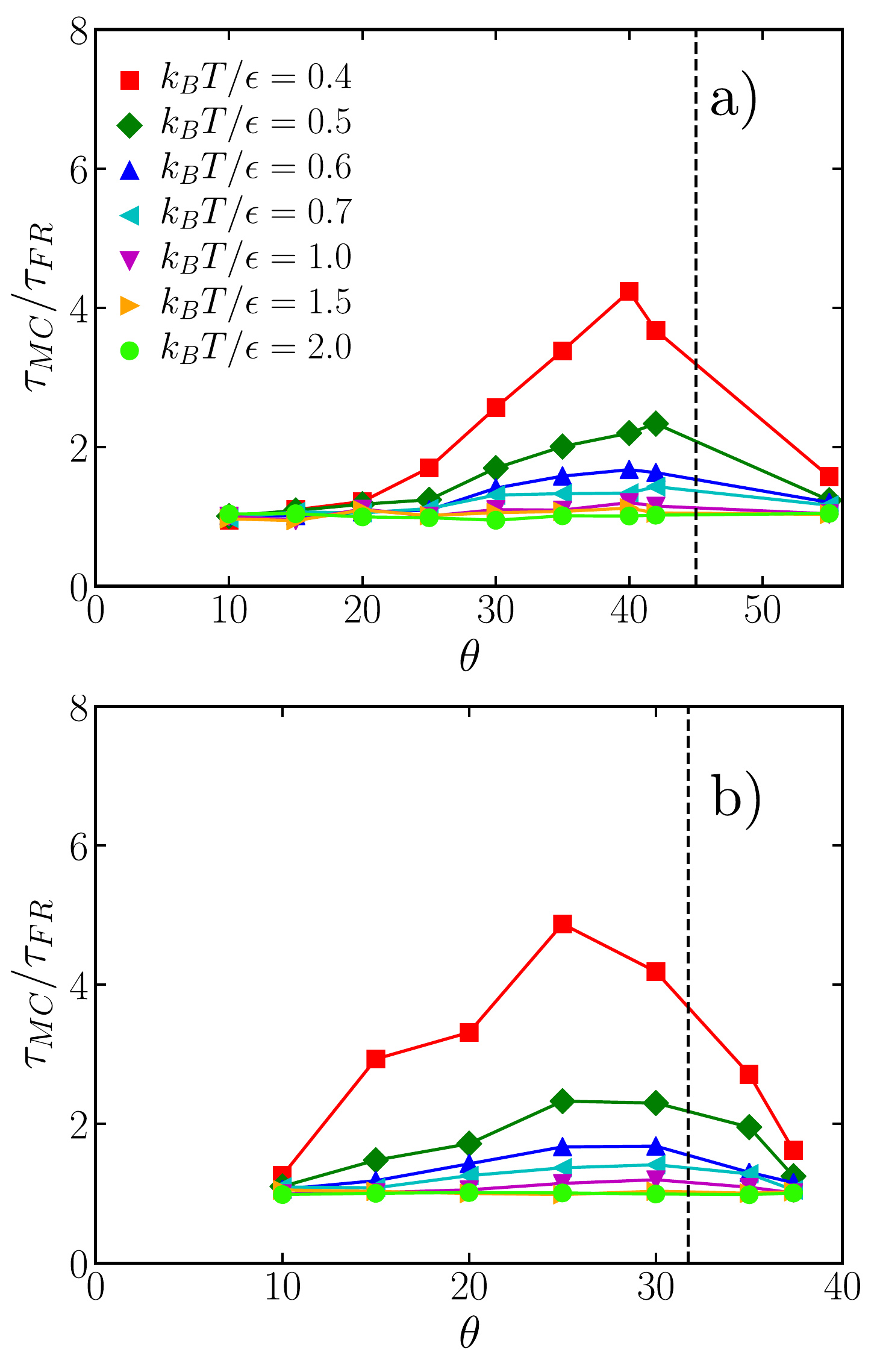}
\end{center}
\caption{Rotational relaxation time $\tau_{DMC}$ obtained from local cage dynamical Monte Carlo simulations. All relaxation times are normalized by the value $\tau_{FR}$ for a free rotator in a simulation with the same step size. Data are shown for both a) 6 patches and b) 12 patches.}
\label{fig:MC}
\end{figure}

In Fig.~\ref{fig:MC} we show the local  dynamical Monte Carlo rotational relaxation time $\tau_{DMC}$ divided by the corresponding relaxation time  $\tau_{FR}$ of a free rotor as function of the angle $\theta$. Comparing the results with Fig.~\ref{fig:tau_angle} we see that the qualitative behavior of the rotations is reproduced by the local DMC approach. In the 6-patch case the relaxation time increases until its maximum shortly before the overlap angle, and then decreases rapidly. In contrast, for the 12-patch case the maximum in the relaxation time shifts to lower angles, ending up around $\theta\simeq 25^\circ$ for the lowest temperatures. Note that in the DMC simulations, the particles behave diffusively at short time scales, as opposed to the ballistic behavior that occurs in EDMD for short times. Hence, the DMC  simulations have significantly longer rotational relaxation times at high temperatures, resulting in a more limited variation in $\tau_{DMC}$ as compared to the $\tau_R$ measured in EDMD. Nonetheless, these results show that the local DMC simulations capture the qualitative behavior of the rotational dynamics in our molecular dynamics simulations. In combination with the large discrepancy in time scales between the rotational and translational dynamics, this confirms that the rotational relaxation of the system is largely decoupled from the translations. Rather than being controlled by global rearrangements, rotational diffusion is dominated by local rearrangements within the translational cage that surrounds a particle.

\subsection{Structural Analysis}

For the 12-patch system, both the EDMD and DMC simulations show a local maximum in relaxation time for relatively small patch sizes. To examine the origin of this feature, we focus our attention on the local structure of the 12-patch systems.

\begin{figure}[!ht]
\begin{center}
 \includegraphics[width=0.9\linewidth]{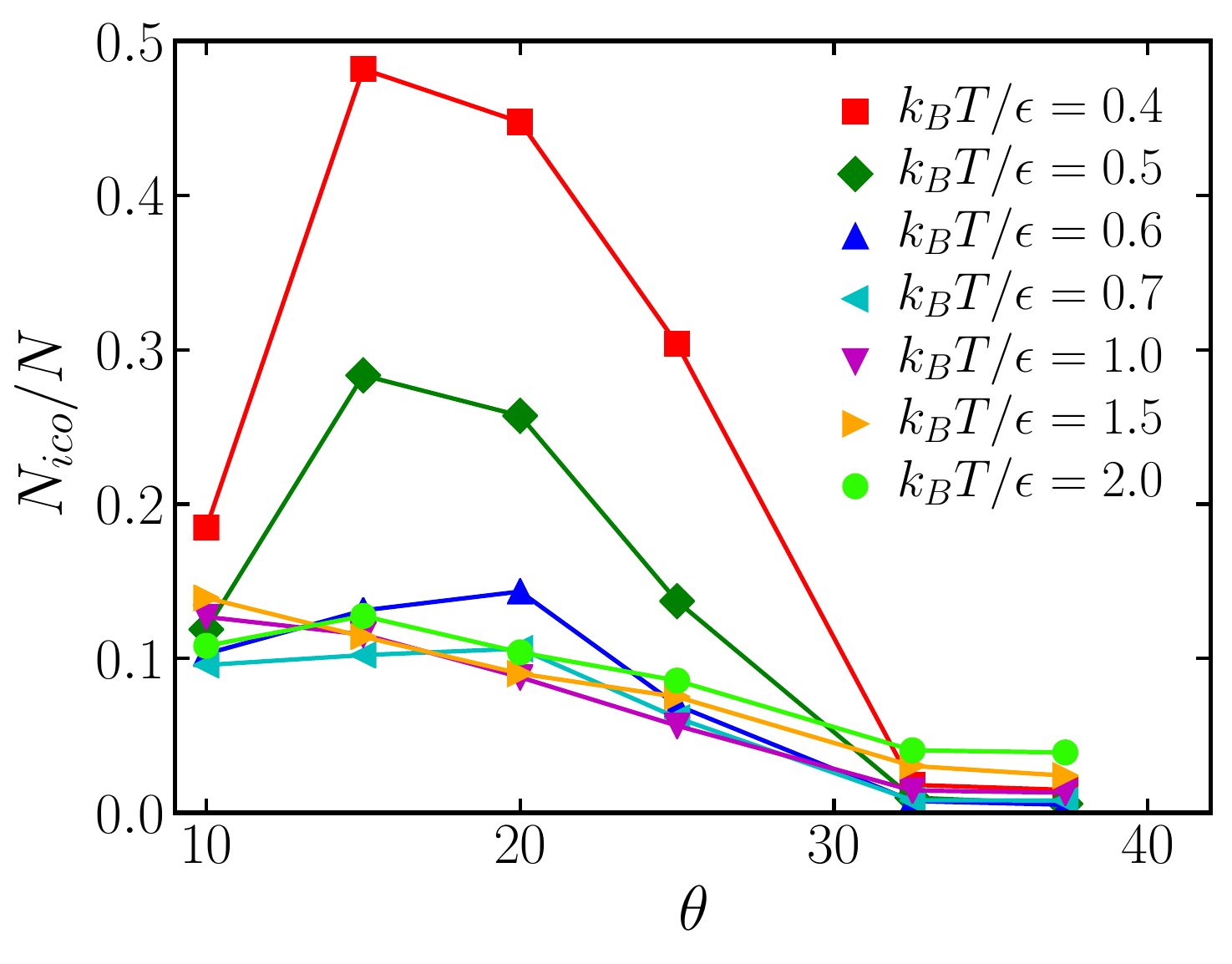} 
\end{center}
\caption{Fraction of particles that are part of icosahedral clusters in the 12-patch system, for different patch sizes $\theta$ and reduced temperatures $k_B T / \epsilon$.}
\label{fig:12ico}
\end{figure}

We use the \textit{Topological Cluster Classification} algorithm \cite{malins2013tcc} to explore the presence of local icosahedral order in our systems. One of the key features of the 12-patch system is its ability to increase the number of local icosahedral clusters in the system \cite{shortpaper}, inducing a dramatic slowdown in the translational dynamics. Hence, these same local structures may be able to explain the behavior of the rotational dynamics as well. In Fig.~\ref{fig:12ico} we show the fraction of particles in the system which are part of an icosahedral environment, as a function of $\theta$. We see a strong maximum in the number of icosahedral structures around a patch angle of $\theta\simeq15^\circ$ at low temperatures. In this regime, a larger fraction of the particles will be enclosed by an icosahedral cage, offering the possibility to bond to a large number of neighbors. This significantly slows down both rotational and translational relaxation, consistent with Figs.~\ref{fig:tau-T} and~\ref{fig:tau_angle}.
 
\begin{figure}[!ht]
\begin{center}
 \includegraphics[width=0.9\linewidth]{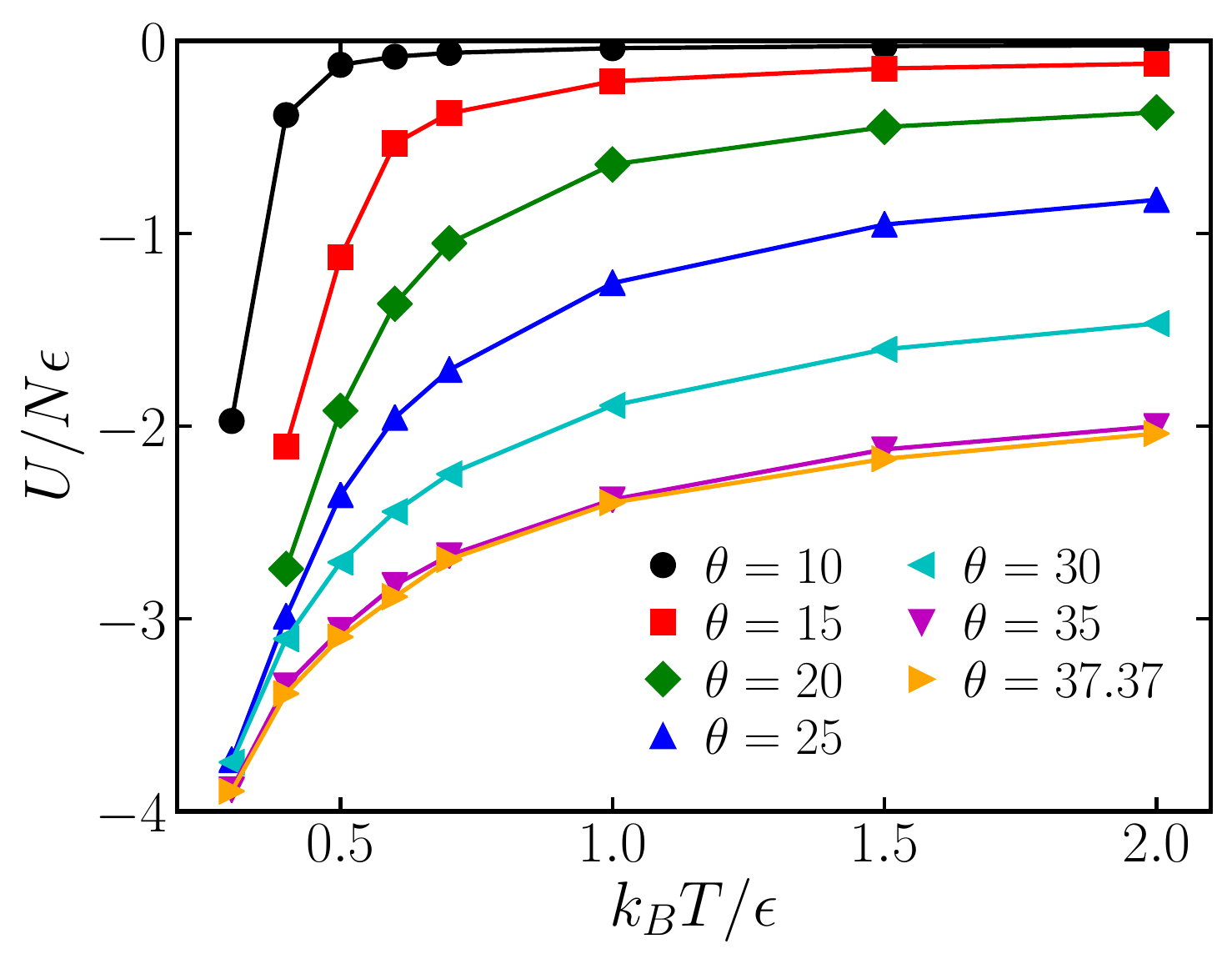} 
\end{center}
\caption{Energy per particle for 12-patch systems with different patch sizes $\theta$. Note that for a fully bonded particle $U / N \epsilon = -6$.}
\label{fig:12energy}
\end{figure}

It is important to note that cage structure is not the only factor determining the rate of rotational relaxation, as the number of bonds formed by the particles also plays an important role. After all, for any given cage structure, particles bound by a larger number of attractive bonds are expected to rotate more slowly. In Fig.~\ref{fig:12energy} we show the energy per particle of the 12-patch case. As one might expect, larger patches lead to a larger number of bonds in the system. Hence, while around $\theta = 15^\circ$ more particles can be found in a (potentially highly bonded) icosahedral environment, the average particle has more bonds when $\theta$ is higher. The rotational relaxation time thus stems from a combination of factors, including the fraction of icosahedral cages (which is maximized for relatively small angles), and the number of bonds (which is maximized for larger angles). Together, these factors contribute to the peak in $\tau_R$ shown in Fig.~\ref{fig:tau_angle}b. 

\section{Conclusions}

The relationship between rotational and translational dynamics in supercooled  liquids is highly complex \cite{dzugutov2002decoupling, edmond2012decoupling} and dependent on the details of the interactions between the particles, including their shape  \cite{zou2019coupling, roos2016coupling, chong2002structural}. Here, we have explored the interplay between local structure, rotational relaxation, and translational relaxation in dense fluids of spherical patchy particles.
The tangled dynamics of the patchy particle systems are also reflected in the relation between translational and rotational dynamics. The rotational relaxation is much faster than the translational relaxation, such that the two relaxations are effectively uncoupled.  Although the changes in the local structure at low temperatures  dramatically slow down both the translational and orientational degrees of freedom, the effect is not sufficiently strong to fully couple the two types of relaxation. Furthermore, the reentrance in the translational dynamics leads to a full decoupling of the rotations from the translations at high temperatures, where the particles can rotate freely but translations are glassy. As a result, for the temperature range explored here, the translational relaxation of the system is determined by slow events that break up the cages in the system, while the particle orientations can relax by simply adjusting their orientation within the existing local cage. This picture is confirmed by our fully local DMC model, which qualitatively reproduces the  rotational behavior of the full system. 

The decoupling between the two types of motion indicates a breakdown of the Stokes-Einstein and Stokes-Einstein-Debye relations. Moreover, the fact that the particles can rotate on time scales much shorter than the translational relaxation time suggests that the patchy interactions in our model can be interpreted as an effective field which manipulates the local structures favored by the system, reminiscent of the biasing potentials used in e.g. Ref. \onlinecite{royall2015role}. Essentially, the main effect of the directional interactions is to enhance or disrupt different types of caging in the system, with dramatic impact on the dynamics of the system (as well as on crystal stability \cite{royall2015strong, shortpaper}). 

In line with earlier work \cite{frank1952supercooling,royall2015role,tarjus2005frustration, royall2018local,miracle2007structural, doye2003favored, dasgupta2019softness}, in the systems studied here icosahedral order is a strong driving force for dynamical arrest. The slowest dynamics, both in terms of translations and rotations, are observed when the system forms a large number of icosahedral clusters. An important observation is the existence of an optimal patch size ($\theta \simeq 15^\circ$) which enhances icosahedral order. Indeed, this optimal patch size is close to the case where we observe the slowest translational relaxation. Hence, tuning patch sizes provides an additional route to manipulating the dynamics of glassy fluids by tuning their local structure.

Our results show that carefully designed particle interactions provide control over the structure and dynamics of dense fluids. While colloidal particles with tunable patchy interactions are now realizable experimentally \cite{choueiri2016surface, yi2013recent,meester2016colloidal, biffi2015equilibrium, smalyukh2018liquid,  bianchi2011patchy}, it is interesting to ask the question whether this strategy extends to models with simpler interactions. For example, tuning the shape of hard anisotropic colloids has been demonstrated to strongly impact the structure of their fluids, as the particle shape gives rise to an effective ``patchiness'' \cite{damasceno2011crystalline}. Indeed, many anisotropic shapes have been found to reliably form glasses\cite{teich2019identity}, while others crystallize \cite{damasceno2012predictive}. Hence, extending this research to anisotropically shaped particles is an extremely promising future avenue for further exploration of the interplay between local structure and dynamics.

\section{Supplementary Material}
See supplementary material for a further analysis on the rotational correlators.

\section{Acknowledgements}
S. Mar\'in-Aguilar acknowledges CONACyT for funding.

\bibliographystyle{apsrev4-1}
\bibliography{references_paper}
\end{document}